\documentclass[12pt]{article}
\usepackage{epsfig}
\usepackage{amssymb}
\usepackage{amsmath,amsfonts,amssymb,graphicx}
\newcommand {\be}{\begin{equation}}
\newcommand {\ee}{\end {equation}}
\newcommand{\beq}{\begin{eqnarray}}
\newcommand{\eeq}{\end{eqnarray}}

\begin{document}
\vspace{-2cm}

\title{Sterile Plus Active Neutrinos and Neutrino Oscillations}
\author{Leonard S. Kisslinger\\
Department of Physics, Carnegie Mellon University, Pittsburgh, 
PA 15213}
\maketitle
\date{}
\noindent
PACS Indices:11.30.Er,14.60.Lm,13.15.+g\vspace{0.25 in}
\begin{abstract}

Using a 3 + 1 neutrino model with one sterile and the three standard active
neutrinos with a 4x4 unitary transformation matrix, U, relating flavor to 
mass neutrino states, the 
probability of $\nu_\mu$ to  $\nu_e$ transition is estimated using 
sterile-active neutrino masses determined by MiniBooNE and other experiments 
and sterile-active neutrino angles in the 4x4 U matrix.
\end{abstract}
\section{Introduction}

  This is an extension of work on time reversal violation\cite{hjk11} and
CP violation\cite{khj12} via neutrino oscillations. That work used S-matrix
theory with a 3x3 matrix to relate the standard three flavor neutrinos to
three neutrinos with well-defined mass. In the present work we extend the
standard model by including a fourth neutrino, a sterile neutrino.
Recent experiments on neutrino oscillations\cite{mini13} (see Ref\cite{mini13} 
for references to earlier experiments) have suggested the 
existence of at least one sterile neutrino and the mass used in the present 
work. See Ref\cite{kmms13} for a discussion
of sterile neutrino oscillations and references to experimental and theoretical
publications, and Ref\cite{tg07} for a 6x6 matrix model. 
Also, pulsar velocities have been estimated using sterile neutrino
emission\cite{kj12} based on a recent estimate\cite{aba12} of the $\nu_e-\nu_s$ 
mixing angle.

Our present work is most closely related to Ref\cite{khj12} in which CPV
was studied. $\mathcal{P}(\nu_a \rightarrow \nu_b)$ is the transition 
probability 
for a neutrino of flavor $a$ to convert to a neutrino of flavor $b$; and 
similarly for antineutrinos $\bar{\nu}_a,\bar{\nu}_b$. The CPV probability 
difference for $\nu_\mu$ to $\nu_e$ oscillation is defined as
\beq
   \Delta\mathcal{P}^{CP}_{\mu e}&=& \mathcal{P}(\nu_\mu \rightarrow \nu_e)
-\mathcal{P}(\bar{\nu}_\mu \rightarrow \bar{\nu}_e) \; .
\eeq

A main objective of Ref\cite{khj12} was to find the dependence of 
$\Delta\mathcal{P}^{CP}_{\mu e}$ on the parameter $\delta_{CP}$. The antineutrino 
oscillation probability is related to the neutrino
oscillation probability by the neutrino matter potential $V\rightarrow -V$
and  $\delta_{CP}\rightarrow -\delta_{CP}$. Since in the present work we are
only investigating how $\mathcal{P}(\nu_\mu \rightarrow \nu_e)$ is modified
by the introduction of a sterile neutrino we set both $V$ and $\delta_{CP}$
equal to zero. $\mathcal{P}(\nu_\mu \rightarrow \nu_e)$ is not very dependent
on either quantity\cite{khj12}.

\section{$\mathcal{P}(\nu_\mu \rightarrow \nu_e)$ Derived Using
the U Matrix} 

  This is an exension of the method introduced by Sato and 
collaborators for three active neutrino oscillations\cite{as96,ks99}
to three active neutrinos plus one sterile neutrino. Active neutrinos with
flavors $\nu_e,\nu_\mu,\nu_\tau$ and a sterile neutrino $\nu_s$ are 
related to neutrinos with definite mass by
\beq
\label{f-mrelation}
      \nu_f &=& U\nu_m \; ,
\eeq
where $U$ is a 4x4 matrix and $\nu_f,\nu_m$ are 4x1 column vectors, which
is an extension of the 3x3 matrix used in Refs.\cite{as96,ks99} (with $s_{ij},
c_{ij}=sin\theta_{ij},cos\theta_{ij}$),
\beq
\label{Uform}
     U &=& O^{23}\phi O^{13} O^{12} O^{14} O^{24} O^{34} {\rm \;\;with}
\eeq
$O^{23}$=
 \( \begin{array}{clcr} 1 & 0 & 0 & 0 \\ 0 & c_{23} & s_{23} & 0 \\
0 & -s_{23} & c_{23} & 0 \\ 0 & 0 & 0 & 1  \end{array} \) 
\vspace{5mm}
$O^{13}$=
 \( \begin{array}{clcr} c_{13} & 0 & s_{13} & 0 \\ 0 & 1 & 0 & 0 \\
-s_{13} & 0  & c_{13} & 0 \\ 0 & 0 & 0 & 1  \end{array} \)
$O^{12}$=
 \( \begin{array}{clcr} c_{12} & s_{12} & 0 & 0 \\ -s_{12} & c_{12} & 0 & 0 \\
 0 & 0  & 1 & 0 \\ 0 & 0 & 0 & 1  \end{array} \)

$O^{14}$=
 \( \begin{array}{clcr} c_\alpha & 0 & 0 & s_\alpha \\ 0 & 1 & 0 & 0 \\
 0 & 0  & 1 & 0 \\ -s_\alpha & 0 & 0 & c_\alpha  \end{array} \)
$O^{24}$=
 \( \begin{array}{clcr} 1 & 0 & 0 & 0 \\ 0 & c_\alpha & 0 & s_\alpha \\
 0 & 0  & 1 & 0 \\ 0 & -s_\alpha & 0 & c_\alpha  \end{array} \)
$O^{34}$=
 \( \begin{array}{clcr} 1 & 0 & 0 & 0 \\ 0 & 1 & 0 & 0 \\
 0 & 0  & c_\alpha & s_\alpha \\ 0 & 0 & -s_\alpha & c_\alpha  \end{array} \)

$\phi$=
 \( \begin{array}{clcr} 1 & 0 & 0 & 0 \\ 0 & 1 & 0 & 0 \\
 0 & 0  & e^{i\delta_{CP}} & 0 \\ 0 & 0 & 0 & 1  \end{array} \)
\vspace{5mm}

\noindent
with $c_{12}=.83,\;s_{12}=.56,\;s_{23}=c_{23}=.7071$. We use 
$s_{13}=.15$ from the recent Daya Bay Colaboration\cite{DB3-7-12}.
In our present work we assume the angles $\theta_{j 4} \equiv \alpha $ for
all three j, and $s_\alpha,c_\alpha=sin\alpha,cos\alpha$. An important aspect of 
our work is to find the dependence of neutrino oscillation probabilities
on $s_\alpha,c_\alpha$.

From Eq(\ref{Uform}) the 4x4 $U$ matrix is
\vspace{2mm}
\small

 \( \begin{array}{clcr} c_{12}c_{13}c_\alpha &c_{13}(s_{12}c_\alpha-c_{12}s_\alpha^2)& 
- c_{13}s_\alpha^2(c_{12}c_\alpha+s_{12})+s_{13}c_\alpha&  c_{13}s_\alpha c_\alpha
(c_{12}c_\alpha+s_{12})+s_{13}s_\alpha \\ A c_\alpha& -As_\alpha^2+Bc_\alpha &
 -As_\alpha^2c_\alpha-Bs_\alpha^2+c_{13}s_{23}e^{i\delta_{CP}}c_\alpha &
As_\alpha c_\alpha^2+Bs_\alpha c_\alpha+c_{13}s_{23}e^{i\delta_{CP}}s_\alpha\\
Cc_\alpha & -Cs_\alpha^2+Dc_\alpha & -Cs_\alpha^2c_\alpha-Ds_\alpha^2+
c_{13}c_{23}e^{i\delta_{CP}}c_\alpha &
Cs_\alpha c_\alpha^2+Ds_\alpha c_\alpha+c_{13}s_{23}e^{i\delta_{CP}}s_\alpha\\
-s_\alpha & -s_\alpha c_\alpha & -s_\alpha c_\alpha^2 & c_\alpha^3  \end{array} \)
\vspace{2mm}
\normalsize
with
\vspace{5mm}
\beq
\label{Uparameters}
   A&=& -(c_{23}s_{12}+c_{12}s_{13}s_{23}e^{i\delta_{CP}}) \nonumber \\
   B&=& (c_{23}c_{12}-s_{12}s_{13}s_{23}e^{i\delta_{CP}}) \\
   C&=& (s_{23}s_{12}-c_{12}s_{13}c_{23}e^{i\delta_{CP}}) \nonumber \\
   D&=& -(s_{23}c_{12}+s_{12}s_{13}c_{23}e^{i\delta_{CP}}) \nonumber \; .
\eeq

Using the formalism of Refs.\cite{as96,ks99} extended to four neutrinos,
the transition probability $ \mathcal{P}(\nu_\mu \rightarrow\nu_e)$ is
obtained from the 4x4 U matrix and the neutrino mass differences
$\delta m_{ij}^2=m_i^2-m_j^2$ for a neutrino beam with energy $E$ and baseline
$L$ by\cite{as96}
\beq
\label{Pue-1}
 \mathcal{P}(\nu_\mu \rightarrow\nu_e) &=& \sum_{i=1}^{4}\sum_{j=1}^{4}
U_{1i}U^*_{1j}U^*_{2i}U_{2j} e^{-i(\delta m_{ij}^2/E)L} \; ,
\eeq
or, since with $\delta_{CP}=0$ $U^*_{ij}=U_{ij}$,
\beq
\label{Pue-2}
\mathcal{P}(\nu_\mu \rightarrow\nu_e) &=& U_{11}U_{21}[ U_{11}U_{21}+
 U_{12}U_{22} e^{-i\delta L}+ U_{13}U_{23} e^{-i\Delta L}+
U_{14}U_{24} e^{-i\gamma L}]+ \nonumber \\
  &&  U_{12}U_{22}[ U_{11}U_{21}e^{-i\delta L}+ U_{12}U_{22} + U_{13}U_{23} 
e^{-i\Delta L}+U_{14}U_{24} e^{-i\gamma L}]+ \nonumber \\
  &&  U_{13}U_{23}[ U_{11}U_{21}e^{-i\Delta L}+ U_{12}U_{22}e^{-i\Delta L}
 + U_{13}U_{23} +U_{14}U_{24} e^{-i\gamma L}]+ \nonumber \\
   &&  U_{14}U_{24}[(U_{11}U_{21}+ U_{12}U_{22} + U_{13}U_{23})e^{-i\gamma L}
 +U_{14}U_{24}] \; ,
\eeq 
with $\delta=\delta m_{12}^2/2E,\; \Delta=\delta m_{13}^2/2E,\; \gamma=
\delta m_{j4}^2/2E$ (j=1,2,3). The neutrino mass differences are 
$\delta m_{12}^2=7.6 \times 10^{-5}(eV)^2$, $\delta m_{13}^2 = 2.4\times 10^{-3} 
(eV)^2$; and we use both $\delta m_{j4}^2=0.9 (eV)^2$ and 
$\delta m_{j4}^2=0.043 (eV)^2$, since $\delta m_{j4}^2=0.043 (eV)^2$ was the
best fit parameter found via the 2013 MiniBooNE analysis, while
$\delta m_{j4}^2=0.9 (eV)^2$ is the best fit using the 2013 MiniBooNE data
and previous experimental fits\cite{mini13}.

  Note that in Refs\cite{hjk11,khj12} $\mathcal{P}(\nu_\mu \rightarrow\nu_e)=
|S_{12}|^2$, with $S_{12}$ obtained from the 3x3 U-matrix and the $\delta m_{ij}$
parameters. Therefore our formalism, given by Eq(\ref{Pue-1}), is quite 
different, and as will be shown for the same $L,E$ the magnitude of
$\mathcal{P}(\nu_\mu \rightarrow\nu_e)$ is also different.
\newpage

From Eq(\ref{Pue-2}),
\beq
\label{Pue}
 \mathcal{P}(\nu_\mu \rightarrow\nu_e) &=& U_{11}^2 U_{21}^2+
 U_{12}^2 U_{22}^2+ U_{13}^2 U_{23}^2+  \nonumber \\
  && U_{14}^2 U_{24}^2+ 2U_{11} U_{21} U_{12} U_{22} cos\delta L + \\
  && 2(U_{11} U_{21} U_{13} U_{22}+ U_{12} U_{22} U_{13} U_{23})cos\Delta L+
\nonumber \\
  &&2U_{14}U_{24}(U_{11} U_{21}+U_{12} U_{22}+U_{13} U_{23})cos\gamma L \; .
\eeq 

Using the parameters given above,
\beq
\label{Uij}
   U_{11}&=& .822 c_\alpha {\rm \;\;\;}U_{12}=-.554s_\alpha^2 +0.084c_\alpha 
\nonumber \\
   U_{13}&=&-.822s_\alpha^2c_\alpha-.554s_\alpha^2+.15c_\alpha {\rm \;\;\;}
U_{14}=.822s_\alpha c_\alpha^2+.554s_\alpha c_\alpha+.15s_\alpha \nonumber \\
   U_{21}&=& -.484c_\alpha {\rm \;\;\;} U_{22}=.484s_\alpha^2 +
.527 c_\alpha \\
    U_{23}&=& .484c_\alpha-.527s_\alpha^2+.7c_\alpha {\rm \;\;\;}
U_{24}=-.484 s_\alpha c_\alpha^2+.527s_\alpha c_\alpha+.7s_\alpha \nonumber \; .
\eeq

\subsection{$\mathcal{P}(\nu_\mu \rightarrow \nu_e)$ for the 3x3 theory}

First we give the results from using the 3x3 theory\cite{khj12} for 
$\mathcal{P}(\nu_\mu \rightarrow \nu_e)$ for comparison with the 4x4 theory 
results given in the next subsection. In this previous work a main goal was 
to study the dependence of $\mathcal{P}(\nu_\mu \rightarrow \nu_e)$ on $s_{13}$,
 but now it has been determined\cite{DB3-7-12} to be approximately 0.15.

The results using the 3x3 theory from Ref\cite{khj12} are shown in Figure 1.

\subsection{$\mathcal{P}(\nu_\mu \rightarrow \nu_e)$ for the 4x4 theory}

With the addition of a sterile neutrino, the 4th neutrino, there are three
new angles, $\theta_{14}$, $\theta_{24}$, and $\theta_{34}$. Our main
assumption is that these three angles are the same, $\theta_{j4}=\alpha$.
The angle $\alpha$ is the main parameter that we are studying.

We also use two values for the sterile-active mass differences. The 
most widely accepted value for $m_4^2-m_1^2$ is $0.9 (eV)^2$\cite{mini13},
but we also use $m_4^2-m_1^2=.043 (eV)^2$ from the 2013 MiniBoonE result
to test the sensitivity of $\mathcal{P}(\nu_\mu \rightarrow \nu_e)$ to
the sterile neutrino-active neutrinos mass differences. Since 
$m_4^2-m_1^2>>m_j^2-m_i^2$ for (i,j)=1,2,3, we assume that $m_4^2-m_j^2=
m_4^2-m_1^2$.

Figure 2 shows our results for $\mathcal{P}(\nu_\mu \rightarrow \nu_e)$
for the four experiments with $m_4^2-m_1^2=0.9 (eV)^2$ and $\alpha=45^o,
60^o,30^o$. As one can see, $\mathcal{P}(\nu_\mu \rightarrow \nu_e)$ is very 
strongly dependent on $\alpha$. 

Next we use $m_4^2-m_1^2=0.043 (eV)^2$, as found in the recent MiniBooNE
experiment, to study the effects of $m_4^2-m_1^2$ on 
$\mathcal{P}(\nu_\mu \rightarrow \nu_e)$, with $\alpha=45^o,30^o,60^o$, as 
shown in Figure 3. 

Note for $\alpha=0$ (no sterile-active mixing) $U_{14}=0$. Therefore,
$\mathcal{P}(\nu_\mu \rightarrow \nu_e)$ is a 3x3 theory;
however, we find that $\mathcal{P}(\nu_\mu \rightarrow \nu_e)$ is different 
with the model of Refs.\cite{as96,ks99}, Eq(\ref{Pue-2}), than the theory 
used in Ref.\cite{khj12}, shown in Figure 1.

\clearpage

\begin{figure}[ht]
\begin{center}
\epsfig{file=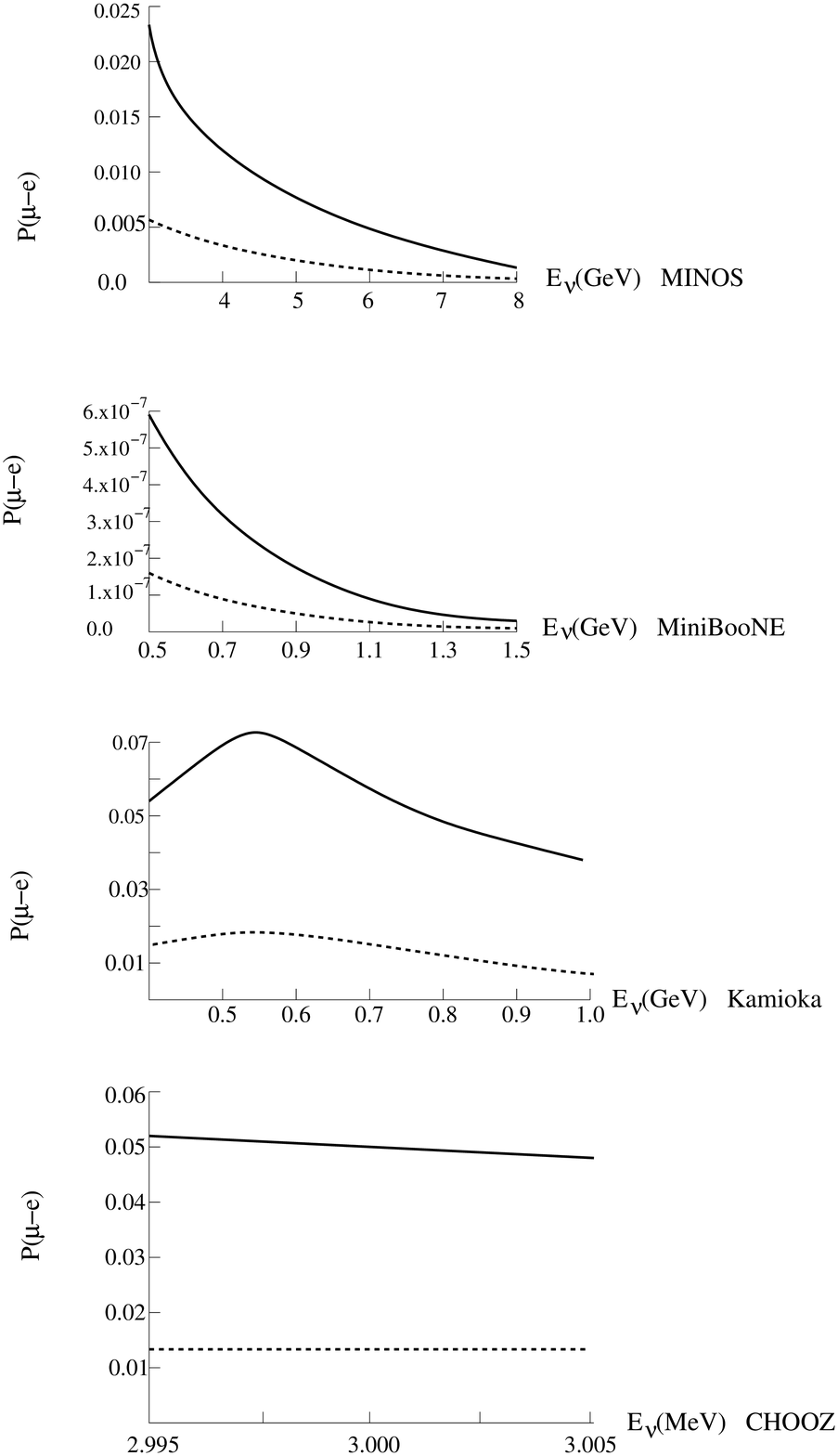,height=16cm,width=12cm}
\end{center}
\caption{\hspace{5mm} The ordinate is $\mathcal{P}(\nu_\mu \rightarrow\nu_e)$ 
for MINOS(L=735 km),
 MiniBooNE(L=500m), JHF-Kamioka(L=295 km), and 
CHOOZ(L=1.03 km) using the 3x3 U matrix.
\hspace {5mm} Solid curve for $s_{13}$=.19 and dashed curve for 
$s_{13}$=.095. The curves are almost independent of $\delta_{CP}$.}
\end{figure}

\clearpage

\begin{figure}[ht]
\begin{center}
\epsfig{file=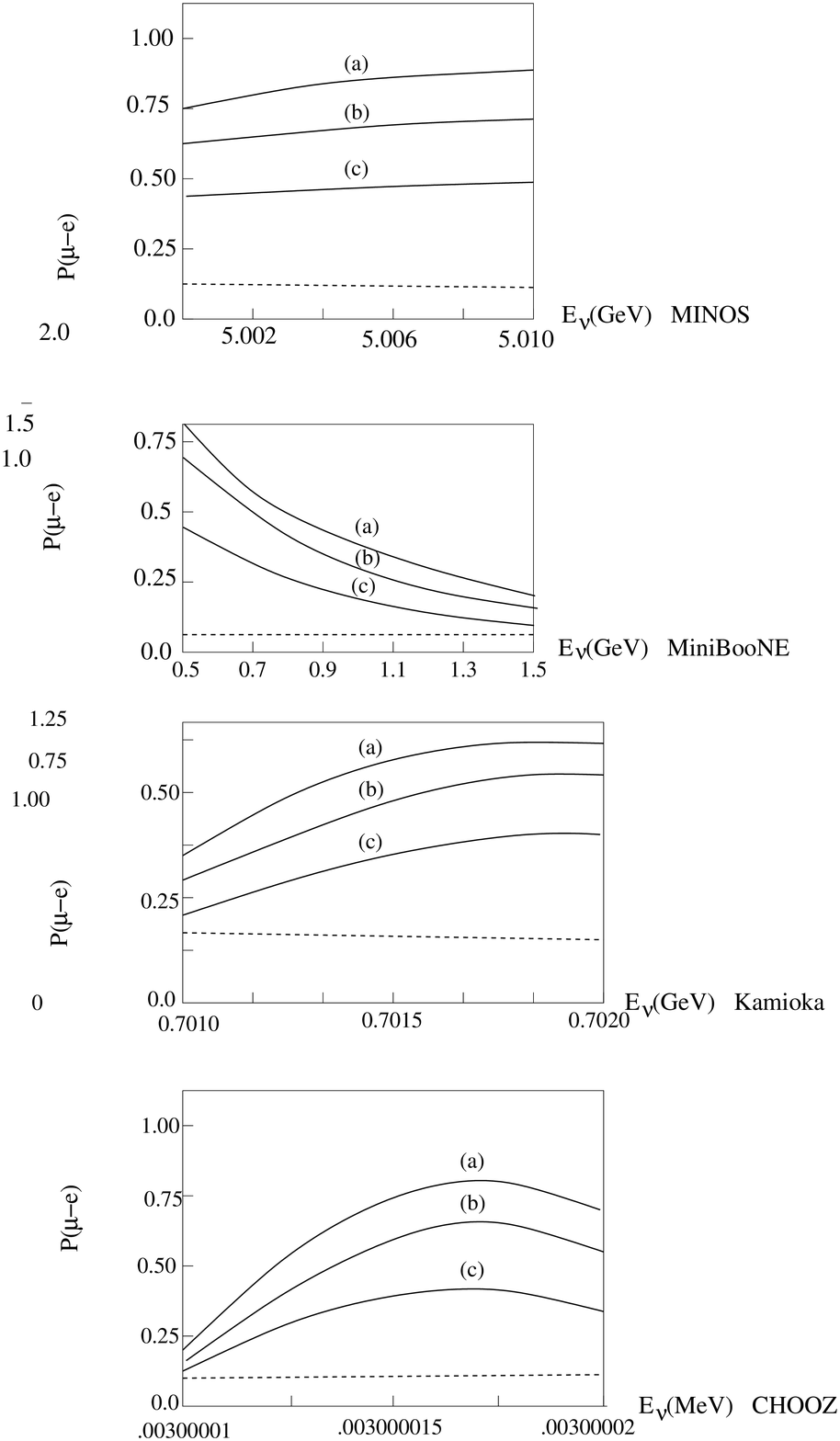,height=16cm,width=12cm}
\end{center}
\caption{\hspace{5mm} The ordinate is $\mathcal{P}(\nu_\mu \rightarrow\nu_e)$ 
for MINOS(L=735 km), MiniBooNE(L=500m), JHF-Kamioka(L=295 km), and 
CHOOZ(L=1.03 km) using the 4x4 U matrix with 
$\delta m_{j4}^2=0.9 (eV)^2$ and (a),(b),(c) for $\alpha=45^o,60^o,30^o$. The 
dashed curves are for $\alpha=0$ (3x3)}
\end{figure}

\clearpage

\begin{figure}[ht]
\begin{center}
\epsfig{file=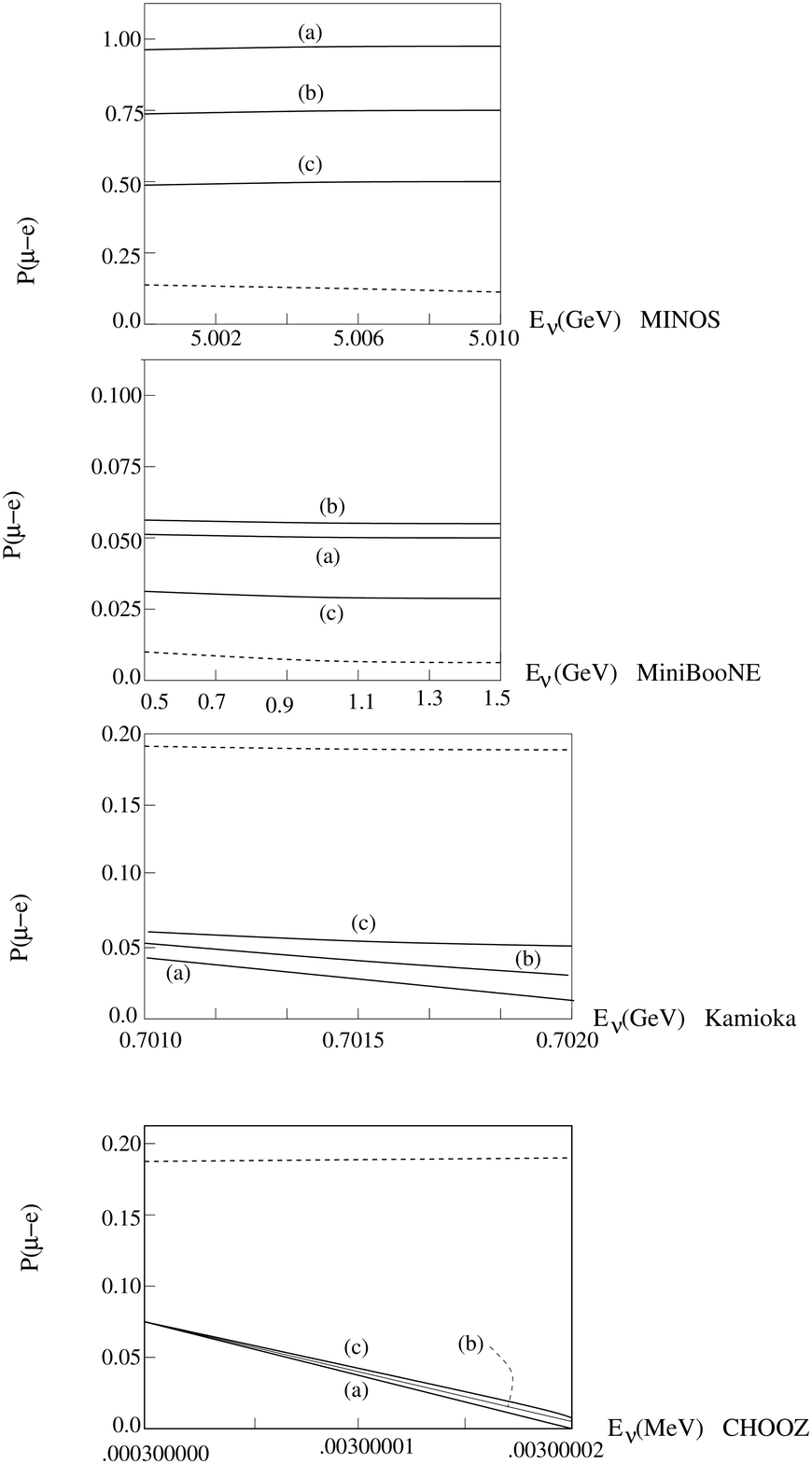,height=16cm,width=12cm}
\end{center}
\caption{\hspace{5mm} The ordinate is $\mathcal{P}(\nu_\mu \rightarrow\nu_e)$ 
for MINOS(L=735 km), MiniBooNE(L=500m), JHF-Kamioka(L=295 km), and 
CHOOZ(L=1.03 km) using the 4x4 U matrix with $\delta m_{j4}^2=0.043 (eV)^2$
and (a),(b),(c) for $\alpha=45^o,60^o,30^o$. The dashed curves are for 
$\alpha=0$ (3x3)}
\end{figure}

\clearpage

\section{Conclusions}

As shown in the figures, the neutrino oscillation probability,
$\mathcal{P}(\nu_\mu \rightarrow \nu_e)$, is quite different for a model
with four neutrinos, $\nu_e,\nu_\mu,\nu_\tau,\nu_s$.
$\mathcal{P}(\nu_\mu \rightarrow \nu_e)$ is also different
for the sterile-active neutrino mass difference $m_4^2-m_1^2=0.043 (eV)^2$ 
vs $m_4^2-m_1^2=0.9 (eV)^2$, which is favored by experiment.

Our most important result is that  $\mathcal{P}(\nu_\mu \rightarrow \nu_e)$ 
is strongly dependent on the sterile-active neutrino mixing angles, with 
the oscillation probability very different for 30, 60 and 45 degrees for
$m_4^2-m_1^2=0.9 (eV)^2$. Therefore, future experiments might be able to 
determine these sterile-active neutrino parameters.  Note that 
Ref.\cite{aba12} favors a small $\nu_e-\nu_s$ mixing angle.

\vspace{3mm}

\Large
{\bf Acknowledgements}
\vspace{3mm}

\normalsize
This work was carried out while the author was a visitor at Los Alamos
National Laboratory, Group P25. The author thanks Dr. William Louis for 
information about recent neutrino oscillation experiments, discussions of 
the theoretical methods used in the present work, and possible future 
experiments.

\end{document}